\documentstyle[twoside,fleqn,espcrc2]{article}
\input epsf

\newcommand{\AmS}{{\protect\the\textfont2
  A\kern-.1667em\lower.5ex\hbox{M}\kern-.125emS}}

\title{Neutron stars accreting the ISM: Are they fast or slow objects ?}

\author{A. Treves\address{II Faculty of Sciences, University of Milano
Via Lucini 3, 22100 Como, Italy}, M. Colpi\address{Dept. of Physics, 
University of Milano, Via Celoria 16, 20133 Milano, Italy}, 
R. Turolla\address{Dept. of Physics, University of Padova, Via Marzolo 8, 
35131 Padova, Italy} and S. Zane\address{Nuclear and Astrophysics Laboratory, 
University of Oxford, Keble Road, Oxford OX1 3RH, UK}}
       
\begin{document}

\begin{abstract}
Old neutron stars (ONSs) which have radiated away their internal 
and rotational energy may still shine if accreting the interstellar
medium. Rather stringent limits from the analysis of 
ROSAT surveys indicate that most optimistic predictions
on ONSs  observability are in excess of a factor as large as $\sim 100$.
Here we explore two  possible evolutionary scenarios
that may account for the paucity of ONSs.
In the first it is assumed that the ONS population is not too fast
($V<100 \ {\rm km\, s}^{-1}$) and that magnetic field decay guides the
evolution. In the second, NSs move with high speed ($V>100$ km s$^{-1}$) 
and preserve their magnetic field at birth. We find that according to the 
former scenario most ONSs are now in the propeller phase, while in the latter 
nearly all ONSs are silent, dead pulsars.

\end{abstract}

\maketitle 

\section {ONSs Basics}

Isolated old neutron stars (ONSs), i.e.  stars which have 
overcome the pulsar phase, are expected to be as many as
$N\sim 10^9$ in the Galaxy.
Ostriker, Rees, \& Silk \cite{ors70} were the first to suggest 
that accretion of the interstellar medium (ISM) may produce enough 
luminosity 
to make the closest stars observable.
Assuming that accretion proceeds at the Bondi rate
\begin{equation}
\label{mdot}
\dot M\simeq 10^{11}n v_{10}^{-3}\, {\rm g\, s^{-1}}\, ,
\end{equation} 
where $v_{10}=(V_{10}^2+c_{s,10}^2)^{1/2}$, $V_{10}$ is the ONS velocity in 
units of 10 km s$^{-1}$, $c_{s}\sim 10 \ {\rm kms}^{-1}$ is the ISM sound 
speed and $n$  the ISM number density in cm$^{-3}$, the total luminosity
is
\begin{equation}
\label{lum}
L\simeq 7\times 10^{31}n v_{10}^{-3}\, {\rm erg\, s^{-1}}\, .
\end{equation} 
In the case of blackbody emission from the polar caps, the emitted radiation 
spectrum peaks around
\begin{equation}
\label{teff}
T_{eff}\simeq 10^{6} L_{31} B_9^{1/7}n^{-1/14} v_{10}^{3/14}\, {\rm K}
\end{equation} 
where $B_9$ is the surface magnetic field in units of $10^9$ G.

Treves, \& Colpi \cite{tc91} and Blaes, \& Madau
\cite{bm93} estimated that, under rather favorable conditions, several 
thousands 
ONSs should appear in the ROSAT All Sky Survey and a few in each deep
field. However, 
despite the intense observational
efforts, the search for ONSs produced, up to now, just a handful of 
candidates, 
out of which only two, RX J18653.5-3754 \cite{wwn96} and  RX J0720.4-3125 
\cite{hab96,hab97}
seem promising. Moreover, recent analyses of ROSAT fields 
in the direction of Giant Molecular Clouds \cite{bzc96,mo97} 
placed rather stringent upper limits 
on the number of ONSs in ROSAT images indicating that current theoretical 
predictions are in excess of at least a factor $\sim$ 5--10. 
A even more dramatic reduction of the number of ONSs could follow from
the recent investigation of ROSAT sources in dark clouds and
high latitude molecular clouds by 
Danner \cite{da96}. For these fields predictions seem 
to be in excess by a factor $\sim 10-100$.

\section{Why Are ONSs  Elusive ?}

Several facts may actually conspire to make detectable ONSs rare. 

\medskip\noindent
a) ONSs could be a faster population than what assumed in 
previous studies which relied on the Narayan \& Ostriker \cite{no90}
velocity
distribution at birth, because either the pulsar 
distribution peaks at higher velocities 
\cite{ll94} or of successive dynamical heating \cite{mb94}. 
The higher the average speed, the lower the luminosity, since it is
$L\propto v^{-3}$. Very recent studies 
\cite{cor97,hp97} seem to confirm that far less pulsars populate
the low--velocity tail $(V<100 \ {\rm km\,s}^{-1})$, relative to Narayan \& 
Ostriker's distribution (see however \cite{hart97}).

\smallskip\noindent
b) The X--ray radiation produced in the accretion process may heat the
surrounding medium \cite{bwm95}. After an initial 
``burst'',
the sound speed in the ISM becomes large enough to cut drastically the 
accretion rate (see eq. [\ref{mdot}]).

\smallskip\noindent
c) The assumptions that ONSs have only a relic field $\sim 10^9$ G
and that {\it all of them} are in the accretion stage now  
oversimplify the picture.
The number of ONSs accreting the ISM may depend sensitively
on the spin and magnetic field evolution. 

\section {Slowly moving ONSs with decaying magnetic field}

Treves, Colpi \& Lipunov \cite{tcl93} and Blaes \& Madau \cite{bm93} 
noted that
for accretion to be possible, the NS must have spun down significantly.
Spin--down is provided first by dipole
losses in the pulsar phase and subsequently by the interaction of the 
rotating magnetosphere with the incoming material in the propeller phase
(see \cite{is75}).
Only when the rotational velocity is low enough to make the centrifugal pull
no longer effective, accretion onto the star surface can occur.
Since the spin--down rates depend on both the star magnetic field and the
period, the ``ability'' to accrete is strongly related to the evolution of 
$B$ and $P$. 
Livio, Xu \& Frank \cite{lxf97} and Colpi, Turolla, Zane \& Treves 
\cite{ctzt97} (CTZT hereafter)
proposed that the deficiency of detected ONSs is related to  
the magnetic field and its decay, and is explained in terms of 
a global reduction in the number of accreting objects. They have
shown that magnetic field decay 
reduces the efficiency of spin--down (both by dipole radiation and by
propeller) to such an extent that the star never enters the accretion stage.
Under rather general assumptions, and regardless of the details of field 
evolution (ohmic decay of a crustal field or spin--induced decay of the core 
field), the number of accretors decreases by more than one order of
magnitude if the characteristic decay time is $\approx 10^8$ yr.
%
In addition a large fraction of ONSs  would be
in the propeller state.
These models are successful in
explaining the spin period ($P=8.39$ s) observed 
in the ONS candidate RXJ0720.4-3125 \cite{hab97,wa97,kp97}. 
CTZT also noticed that 
nearly all ONSs would be accreting today, 
if some physical processes in the crust and/or in the NS interior prevent
$B$ from decaying substantially, and this would contrast with existing
observational limits.

\section {Fastly moving ONSs with non--decaying magnetic field}

Because of the uncertainties in the velocity distribution of 
pulsars, and hence of ONSs, it may be useful to discuss
the status of ONSs observability considering a distribution function
devoided of low--velocity stars. In the following we adopt a different
view for the evolution of the $B$--field, assuming that ONSs retain their
magnetic field at birth. Since no decay on timescales shorter than the 
typical pulsar lifetime ($\approx 10^7$ yr) has been observed so far
and because decay models are far from giving univocal predictions, 
the no--decay scenario is at least as plausible as the opposite one.

Here we consider the evolution of a NS with mass $M=1 M_{\odot}$, 
$R=10$ km and moment of inertia $I=0.8\times 10^{45} \rm{g\, cm}^2$. $B_{12}$
is the constant field in units of $10^{12}$ G, $P$ is the star rotational
period and the notation of CTZT is used.
Over its lifetime, a NS goes through three main evolutionary 
phases which are discussed in some detail below.

\subsection{Emitter phase}

This  regime begins with the pulsar phase and proceeds also after the
break--down of the coherence condition which quenches the radio emission
(the star becomes a dead, or silent, pulsar).
The energy losses are due to magnetic dipole radiation and the period
increases in time according to
\begin{equation}
\label{pdipol}
P=3\times 10^{-4} B_{12} t^{1/2}\, \rm{s}
\end{equation}
where $t$ is in yr.

\subsection{Propeller phase}

When the gravitational energy density of the incoming interstellar gas exceeds
the outward momentum flux of the star at the accretion radius,
$r_{ac}\simeq GM/v^2$, matter starts to accrete. For this condition to be met,
the period must have reached  a critical value
\begin{equation}
\label{ppropel}
P_{prop}\simeq 10 n^{-1/4}v_{10}^{1/2} B_{12}^{1/2}\, \rm {s}
\end{equation}
which is attained by dipole braking in a time 
\begin{equation}
\label{tdipol}
t_{e}\simeq 1.1\times 10^9 n^{-1/2} v_{10} B^{-1}_{12}\,\rm {yr}\,.
\end{equation}
When $P>P_{prop}$  matter can penetrate down to the Alfven 
radius, $r_A\simeq 1.7\times 10^{10}  n^{-2/7} v_{10}^{6/7}B_{12}^{4/7} \,
{\rm cm}$, but the interaction with the rotating magnetosphere prevents 
accretion to go any further because of the centrifugal barrier. The NS is now 
in the propeller phase, rotational energy is lost to the ISM and the period 
keeps increasing at a rate 
\begin{equation}
\label{rateprop}
\frac{dP}{dt}\simeq K\,P^{\alpha} \ \rm{s\, yr}^{-1}
\end{equation}
with $K=\times 10^{-8} n^{9/13} v_{10}^{-27/13} B_{12}^{8/13}$
and $\alpha=21/13$.

\subsection{Accretor phase}

Accretion to the star surface occurs when the corotation radius 
$r_{co}=(GM\,P^2/4\pi^2)^{1/3}$ becomes larger   
than the Alfven radius. This implies that spin--down increased the period up
to
\begin{equation}
\label{paccr}
P_{ac}\simeq 2.5\times 10^3 n^{-3/4} v_{10}^{1/7}B_{12}^{6/7}\, \rm s\,.
\end{equation}
Using equations (\ref{rateprop}) and (\ref{paccr}), the duration of the
propeller phase turns out to be 
\begin{equation}
\label{tprop}
t_{p}={P_{prop}^{1-\alpha}-P_{ac}^{1-\alpha}\over K(\alpha -1)}\, \rm yr\,.
\end{equation}

\begin{figure}[htb]
\vspace{5pt}
\epsfxsize=75mm\epsfysize=65mm\epsfbox{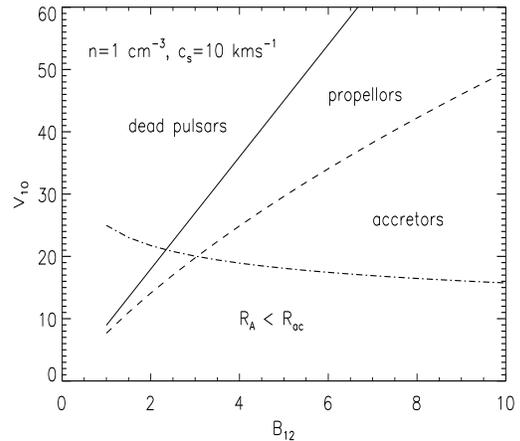}
\caption{The different phases of ONSs at present as a function of the star
velocity and magnetic field.}
\label{fign1}
\end{figure}

\begin{figure}[htb]
\vspace{5pt}
\epsfxsize=75mm\epsfysize=65mm\epsfbox{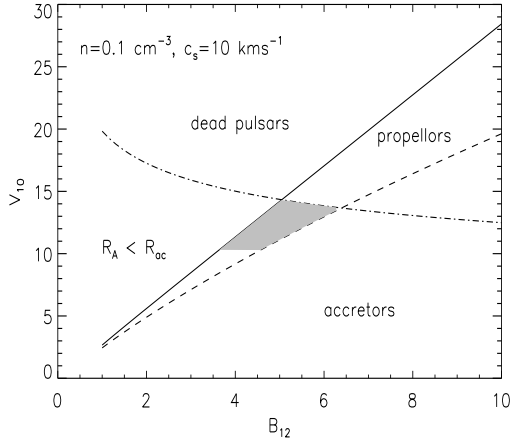}
\caption{Same as in figure \ref{fign1} for $n=0.1 \ {\rm cm}^{-3}$.}
\label{fign01}
\end{figure}

The two relevant timescales $t_e$ and $t_p$, the duration
of the emitter and the propeller phase, depend on the star
velocity, the ISM density and the magnetic field. In order for the star 
to become an accretor, a time $t_a=t_e+t_p$ has to elapse. 
Imposing the condition that each of the times is less (or
equal) to the star lifetime, which we assume to coincide with 
the age of the galaxy $t_G=10^{10}$ yr, gives the velocity range within
which the NS is, at present, a dead pulsar, a propellor or an accretor.
Actually the condition that $P\geq P_{ac}$ [eq.(\ref{paccr})] is not 
sufficient to guarantee that matter is captured at the accretion
radius. At the very low accretion
rates expected for fast, isolated ONSs, it could be that the Alfven radius is
larger that the accretion radius. The condition
$r_A<r_{ac}$ translates into a limit for the star velocity
\begin{equation}
\label{vlim}
v_{10} < 25\, n^{1/10}B_{12}^{-1/5}
\end{equation}
which needs to be taken into account together with $t_p< t_G$ and $t_a<t_G$.
Results are shown in figures \ref{fign1} and \ref{fign01} where the velocities 
at which $t_e=t_G$ (solid line) 
and $t_a=t_G$ (dashed line), together with the limit implied by 
(\ref{vlim}) (dash--dotted line), are plotted 
against the star magnetic field for $n=1 \ {\rm cm}^{-3}$ and $n=0.1 \ 
{\rm cm}^{-3}$.

\section {Discussion}

As can be seen from figure \ref{fign1}, low--velocity ONSs ($V<70 \ {\rm km\, 
s}^{-1}$) moving through the average ISM are in the accretion phase now 
for all reasonable values of the (constant) magnetic field. If their 
velocity distribution at
birth is like that of Narayan \& Ostriker, there is an excess of accretors 
with respect to what implied by ROSAT observations. As discussed in $\S 3$,
a possible way out is to invoke some form of field decay on timescales 
$< 10^9$ yr. In this case, CTZT have found that for spin--induced decay 
the number of  propellers may exceed that of accretors.

On the other hand, if typical ONSs velocities are $>100 \ {\rm km\, s}^{-1}$,
as  favoured by most recent observations of pulsars velocities, most of
them should have moved at relatively high Galactic latitudes, $z\sim 300$ pc
or larger. In this case an appropriate value of the ISM density should be 
$0.1 \ {\rm cm}^{-3}$ (see figure \ref{fign01}). All ONSs with 
$V>150-200 \ {\rm km\, s}^{-1}$ are 
above the limit $r_A = r_{ac}$ and will never be detectable, being unable 
to accrete. Although also fast stars with large enough fields ($B_{12}>5$)
may accrete, their luminosity will be much less than that of slow accretors, 
and the expectation of observing one of these objects is practically nil. 
There is, however, the possibility that the objects which populate the
propeller region (the shaded area in figure 2) may be detectable. Even if
their number is not expected to be large, the duration of the propeller
phase is typically about $10^9$ yr, so there could be the possibility
of catching these sources active.
In fact, as proposed by Treves, Colpi \& Lipunov \cite{tcl93}, 
ONS--propellers may 
exhibit a flaring activity due to  episodic accretion of the matter
which piles up at the Alfven radius. The flaring luminosity
can be  high, so one can hope to detect these transient sources.

\end{document}